\documentclass[1p]{elsarticle}

\usepackage{lineno,hyperref}

\usepackage{graphicx}
\usepackage{amsmath}

\usepackage{siunitx}

\journal{Nucl. Instr. Meth. Phys. Res. A}

\begin{document}

\begin{frontmatter}

\title{RF design of APEX2 two-cell continuous-wave normal conducting photoelectron gun cavity based on multi-objective genetic algorithm}

\author[lbnl]{T. Luo\corref{mycorrespondingauthor}}
\cortext[mycorrespondingauthor]{Corresponding author}
\ead{tluo@lbl.gov}
\author[lbnl,tshinghua]{H. Feng}
\author[lbnl]{D. Filippetto}
\author[lbnl]{M. Johnson}
\author[lbnl]{A. Lambert}
\author[lbnl]{D. Li}
\author[lbnl]{C. Mitchell}
\author[lbnl]{F. Sannibale}
\author[lbnl]{J. Staples}
\author[lbnl]{S. Virostek}
\author[lbnl]{R. Wells}
\address[lbnl]{Lawrence Berkeley National Laboratory, One Cyclotron Road, Berkeley, California 94720, USA}
\address[tshinghua]{Department of Engineering Physics, Tsinghua University, Beijing 100084, China.}

\begin{abstract}
High brightness, high repetition rate electron beams are key components for optimizing the performance of next generation scientific instruments, such as MHz-class X-ray Free Electron Laser (XFEL) and Ultra-fast Electron Diffraction/Microscopy (UED/UEM). In the Advanced Photo-injector EXperiment (APEX) at Berkeley Lab, a photoelectron gun based on a \SI{185.7}{MHz} normal conducting re-entrant RF cavity, has been proven to be a feasible solution to provide high brightness, high repetition rate electron beam for both XFEL and UED/UEM. Based on the success of APEX, a new electron gun system, named APEX2, has been under development to further improve the electron beam brightness. For APEX2, we have designed a new \SI{162.5}{MHz} two-cell photoelectron gun and achieved a significant increase on the cathode launching field and the beam exit energy. For a fixed charge per bunch, these improvements will allow for the emittance reduction and hence to an incresed beam brightness. The design of APEX2 gun cavity is a complex problem with multiple design goals and restrictions, some even competing each other. For a systematic and comprehensive search for the optmized cavity geometry, we have developed and implemented a novel optimization method based on the Multi-Objective Genetic Algorithm (MOGA).

\end{abstract}


\end{frontmatter}


\section{Introduction}

The high brightness, high repetition rate electron beams are key components for several scientific applications, such as X-ray Free Electron Laser (XFEL)~\cite{lcls-ii, exfel} and Ultra-fast Electron Diffraction/Microscopy (UEM/UEM)~\cite{ued-uem} when MHz-class repetition rates are required. Different types of electron guns have been under development for decades, such as the DC gun~\cite{dc-gun-cornell}, the normal conducting RF gun~\cite{nc-gun-lcls} and the superconducting RF gun~\cite{sc-gun-review}, each with its own advantages and challenges. In the Advanced Photo-injector EXperiment (APEX) at Lawrence Berkeley National Laboratory (LBNL), a photoelectron gun based on a normal conducting re-entrant RF cavity operated at Very-High-Frequency (VHF) \SI{185.7}{MHz} (1/7th of \SI{1.3}{GHz}), has been designed, manufactured and commissioned~\cite{apex-commissioning}. APEX has successfully achieved the emittance requirements for the high brightness, high repetition rate XFEL and demonstrated stable and reliable operations required for a user facility. An electron injector almost identical with APEX has been built and delivered by LBNL as the injector for the Linac Coherent Light Source II (LCLS-II).

The advance of XFEL and UED/UEM requires ever-increasing brightness for electron sources. 
Based on the success of APEX, a new project named APEX2~\cite{apex2_fernando} has been initiated at LBNL. APEX2 aims at further extending the performance of normal conducting gun technology by increasing both the cathode launching field $E\textsubscript{cathode}$ and the output energy $V$. A two-cell \SI{162.5}{MHz} (1/8th of \SI{1.3}{GHz}) cavity with re-entrant structure similar to APEX was chosen as the design baseline for APEX2. The first cell, named the gun cell, provides high $E\textsubscript{cathode}$ and the initial acceleration for the electron beam. The following cell, named the 2\textsuperscript{nd} cell, carries out further acceleration for the beam. 

The requirements on the significant increase for both $E\textsubscript{cathode}$ and $V$, along with constraints from power considerations, engineering feasibility and beam dynamics requirements, impose considerable challenges on the APEX2 cavity RF design. A novel method based on Multi-Object Genetic Algorithm (MOGA)~\cite{moga} has been developed and applied to APEX2 design. Integrating the MOGA algorithm, the numerical electromagnetic ($EM$) field solver, and the parallel computing, this method becomes a useful tool for cavity geometry optimization.   

The paper is organized as follows. First we describe the novel RF cavity design method based on MOGA. Then we present the RF design of the two-cell \SI{162.5}{MHz} gun cavity obtained by this new method. The general properties of the re-entrant VHF gun, a preliminary design of an alternative \SI{216.7}{MHz} (1/6th of \SI{1.3}{GHz}) VHF gun, and plans for the future development of the MOGA-based cavity design method are presented in the DISCUSSION section. 

\section{RF cavity design based on MOGA} 
Most real-world engineering problems involve simultaneously optimizing multi-objectives where considerations of trade-offs are important. In recent decades, Genetic Algorithm~\cite{ga} (GA), which is inspired by the evolutionary theory ``survival of the fittest", has became one of the primary tools to solve real-world multi-objective problems. For particle accelerators, MOGA has already been widely applied in the lattice design for photoinjector~\cite{bazarov-gun-moga}, storage rings~\cite{alsu-moga,nsls-ii-moga} and linear accelerators~\cite{lcls-ii-moga,linac-driver-moga} (LINAC). 

The idea of implementing GA into RF cavity geometry design was first proposed in 90s~\cite{ga-95-headway} but without much follow-up for almost two decades. In recent years, with the advance of both GA algorithms and computation capability, as well as the ever-increasing demand on the cavity performance, more efforts have been invested into this approach\cite{ga-shin,ga-kranjcevic,ga-hofler}. In the conventional design method, the designer starts with an overall cavity shape and defines the geometric variables. During the optimization, the cavity geometry is modified by changing one or a few geometric variables at a time. All the variables are scanned iteratively in a ``trail-and-error" approach until a satisfying result is achieved. This method heavily relies on the designer's experience and judgment. The pre-defined shape where the searching process starts can be over-constrained that potentially better solutions are left out. For complicated geometry with dozens of geometric variables, the scanning process can become tedious. When there are competing optimization goals, the rationale to choose one design over the other is not straightforward. By implementing MOGA into RF cavity design, we aim to establish a quantitatively-defined optimization method with more efficient algorithm and more thorough searching process that can lead to better cavity solution not easily achievable by the conventional method. In what follows we describe in detail this MOGA-based cavity design method.

\subsection{Parameterizing Cavity Geometry}

Assuming a perfectly conducting cavity surface, the cavity $EM$ field is determined by its geometry which sets the boundary condition for the Maxwell's equations. The cavity geometry is described by a geometry vector $\mathcal{G} =\{g_1, g_2, ... \} $, where the $g_{n}$ represent the geometric parameters such as segment lenghts, angles, radius of the arcs, etc.. Cavity optimization is carried out by finding a geometry $\mathcal{G}$ that provides the most desired eigenmode solution. 

Critical aspects of the method reside in the definition of $\mathcal{G}$ and in choosing the proper scanning range for each $g_{n}$. They determine how flexible we can vary the cavity geometry during the optimization and how much the computation cost will be.

\subsection{Solving Cavity EM Field}
Once the cavity geometry $\mathcal{G}$ is given, we can calculate the cavity eigenmodes and the corresponding RF properties. Except for a few cases that can be calculated analytically, most cavities' $EM$ fields are solved numerically with software such as CST~\cite{cst}, ANSYS~\cite{ansys}, COMSOL~\cite{comsol} and ACE3P~\cite{ace3p}. In this paper, we use the 2D solver SUPERFISH~\cite{superfish} for its fast speed and built-in post-processing functions. The cavity geometry $\mathcal{G}$ is translated into an input file for SUPERFISH. Then the relevant RF properties are extracted from the SUPERFISH output. Both relevant geometry parameters and RF properties constitute the figure of merit vector for the cavity $\mathcal{M} =\{m_1, m_2,...\}$, where the $m_{n}$ represent the cavity frequency $f$, cathode launching field $E\textsubscript{cathode}$, cavity radius $R$, etc..

\subsection{Implementing MOGA}

As a multi-objective optimization problem, a cavity RF design can be formatted into finding a geometries $\mathcal{G}$ that
\begin{equation}
\begin{split}
\textrm{Minimize}& \;\;\; o_i(\mathcal{G}), i=1,2,...;\\
\textrm{while are subjected to} &\;\;\; c_j(\mathcal{G}) \le 0, j=1,2,...;\\
 & \;\;\;g_n^L<g_n<g_n^U,
\end{split}
\end{equation}
where the objective vector $\mathcal{O} =\{o_1, o_2,...\}$ represents the design goals and the constraint vector  $\mathcal{C} =\{c_1, c_2,...\}$ represents the restrictions, both derived  from cavity's figure of merit vector $\mathcal{M}$.  $g^{L}_{n}$ and $g^U_N$ are respectively the lower and upper limits of the cavity geometry variables $g_n$.
The optimization process results in  a group of geometries $\mathcal{G}_1, \mathcal{G}_2,...$  which are non-dominant~\cite{deb-non-dominant} over each other. They are called Pareto-optimal solutions and together they make up the Pareto front in the objective phase space. A final geometry is chosen from the Pareto front based on further considerations.

Being a population based approach, GA is well suited to solve multi-objective optimization problem. Many MOGA algorithms have been developed and implemented in different applications. In this paper we use Non-dominant Sorting Genetic Algorithm II (NSGA-II)~\cite{nsga-ii} for its well-tested performance and high efficiency. The detail of the cavity optimization process is described as followsing:

\begin{enumerate}
\item{Randomly generate the initial set of cavity geometries $ \{\mathcal{G}_i\}$. }
\item{ Calculate the relavent properties of each geometry by SUPERFISH $ \{\mathcal{G}_i\} \rightarrow \{\mathcal{M}_i\}$.}
\item{Assign ranks and crowd distance to each geometry $\{\mathcal{G}_i\}$ based on the design goals $ \{\mathcal{O}_i\}$ and  restrictions $ \{\mathcal{C}_i\}$, where $\mathcal{O}_i\subset\mathcal{M}_i $ and $\mathcal{C}_i\subset\mathcal{M}_i $. }
\item{Produce new geometries from current ones by selection, crossover and mutation: $ \{\mathcal{G}_i\}\rightarrow  \{\mathcal{H}_i\}$.}
\item{Calculate the relevant properties of new cavity geometries by SUPERFISH $ \{\mathcal{H}_i\} \rightarrow \{\mathcal{N}_i\}$. }
\item{Sort $\{\mathcal{H}_i\}$ based on design goals $ \{\mathcal{O}_i\}$ and restrictions $ \{\mathcal{C}_i\}$ by fast-non-dominated-sort, where $\mathcal{O}_i\subset\mathcal{N}_i $ and $\mathcal{C}_i\subset\mathcal{N}_i $.}
\item{Select new generation of geometries  $\{\mathcal{G}^\prime_i\}$ from combined $\{\mathcal{M}_i,\mathcal{N}_i\}$.}
\item{Go back to Step 2, replace $\{\mathcal{G}^\prime_i\} \rightarrow \{\mathcal{G}_i\}$, keep going until reaching targeted generation.}
\end{enumerate}

MOGA is naturally suitable for parallel computing. We parallelized the program with Multiple Passage Interface and carried out the computation on a 12-core local Windows workstation. For a population of 720, it takes about 48 hours to finish the calculation of 200 generations.

\section{RF design of APEX2 two-cell 162.5 MHz gun cavity}

A \SI{162.5}{MHz} two-cell cavity is chosen as the baseline for APEX2. The choice of the frequency allows compatibility with  other frequencies commonly used in LINAC cavities ( i.e. \SI{325}{MHz} and \SI{650}{MHz} for XFEL). Compared to APEX (\SI{185.7}{MHz}), the lower frequency  also helps reduce the surface resistance and therefore the power density on the cavity inner surfaces. With high $E\textsubscript{cathode}$, the gun cell generates high current, high brightness electron beam while providing an output energy similar to the APEX gun. The 2\textsuperscript{nd} cell provides further acceleration up to \SI{1.5}{MeV}. Both cells are of re-entrant structure similar to APEX. The RF coupling between the two cells is negligible, so they can be powered and tuned separately. 

The RF design is intentionally kept similar to the APEX gun, which has already demonstrated several key operating parameters. The new RF field profile of the APEX2 gun has been applied to the beam dynamics studies~\cite{chad-apex2-emittance} to optimize the injector emittance performance. The cavity design is an iterative process interacting with the beam dynamics study and the engineering evaluation. The design of each cell is described in this section. 

\subsection{The Gun Cell Design}

\begin{figure}[hbt!]
\centering
\includegraphics[width=70mm]{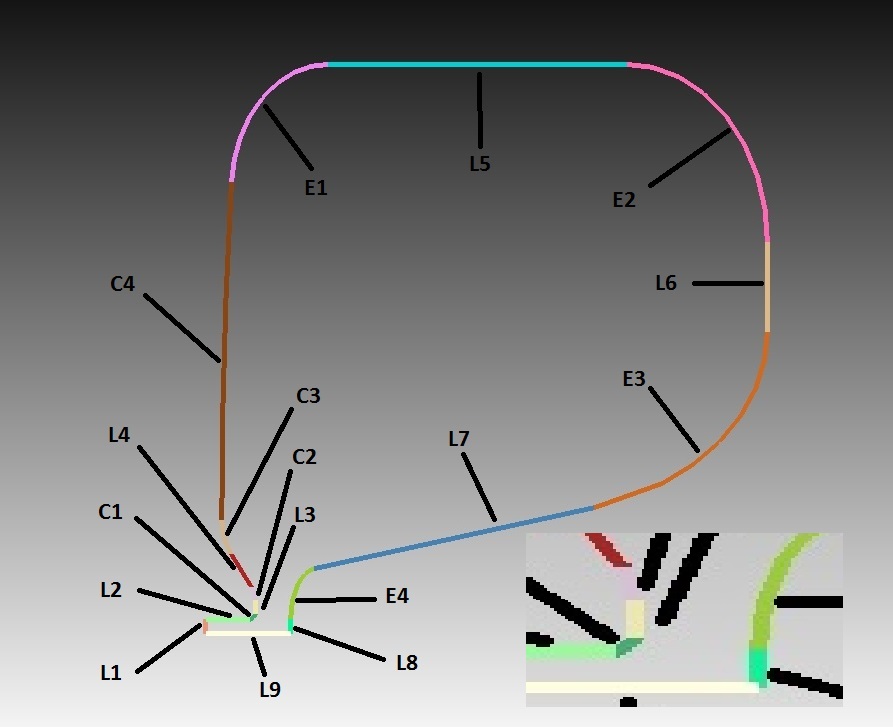}
\caption{\label{fig:gun-cell-outline} Geometrical description of gun cell. It consists nine straight lines L1 to L9, four circular arcs C1 to C4 and four elliptical arcs E1 to E4. The zoom-in view of the gap region is at the bottom right.}
\end{figure}

\begin{figure}[hbt!]
\centering
\includegraphics[width=70mm]{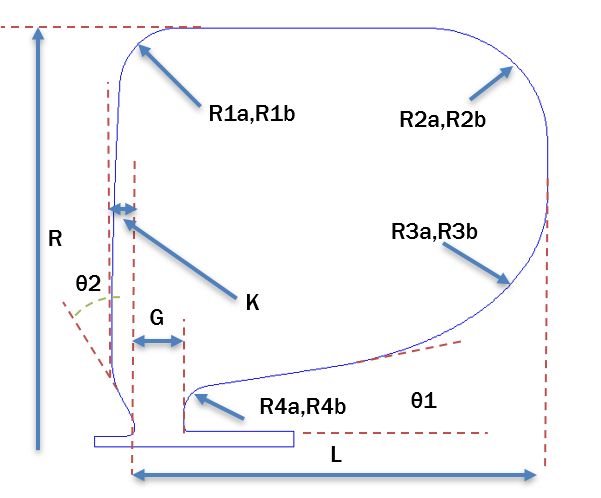}
\caption{\label{fig:geometry} Major geometric parameters of the re-entrant structure cavity. This figure is for the 2\textsuperscript{nd} cell. The gun cell is similar except the cathode plug. }
\end{figure}

The gun cell geometry is described by seventeen segments, including nine straight lines, four circular arcs and four elliptical arcs, as shown in Figure~\ref{fig:gun-cell-outline}. In the simulation, the beampipe radius L1 and cathode flat area L8 are set at \SI{1}{cm}. Beampipe length L2 is set at \SI{10}{cm}. The other 14 segments are described by 19 independent geometric parameters, as shown in Figure~\ref{fig:geometry}. 

The most important optimization goal of the gun cell design is the high launching electric field on the cathode, which determines the beam transverse brightness \cite{FilippettoBrightness}. Also the total RF power should be as small as possible. Other considerations include the RF frequency, peak power density, peak electric field, the practical cavity size limit, the accommodation of the cathode and laser system and so on. These design goals and limits are defined as objectives and constraints in MOGA, as listed in Table~\ref{moga-gun-cell}.

\begin{table*}
\caption{MOGA optimization setting for the gun cell design \label{moga-gun-cell} }
\begin{tabular}{ll}\hline \hline
\textbf{Objectives}&\textbf{Constraints} \\ \hline
With total voltage $V=$ \SI{820}{kV}:  & Peak surface field $E\textsubscript{peak} <$ \SI{37}{MV/m} \\ 
1) Maximize lauching field $E\textsubscript{cathode}$ & Peak power density $PD\textsubscript{peak} <$ \SI{35}{W/cm\textsuperscript{2}} \\
2) Minimize total RF power $P\textsubscript{total}$ & RF frequency $f=162.5 \pm$\SI{3}{MHz} \\
& Cavity radius $R<$ \SI{41}{cm} \\
& Extrusion on anode side $K<$ \SI{2}{cm}\\
\hline\hline
\end{tabular}
\end{table*}

\begin{figure}[hbt!]
\centering
\includegraphics[width=80mm]{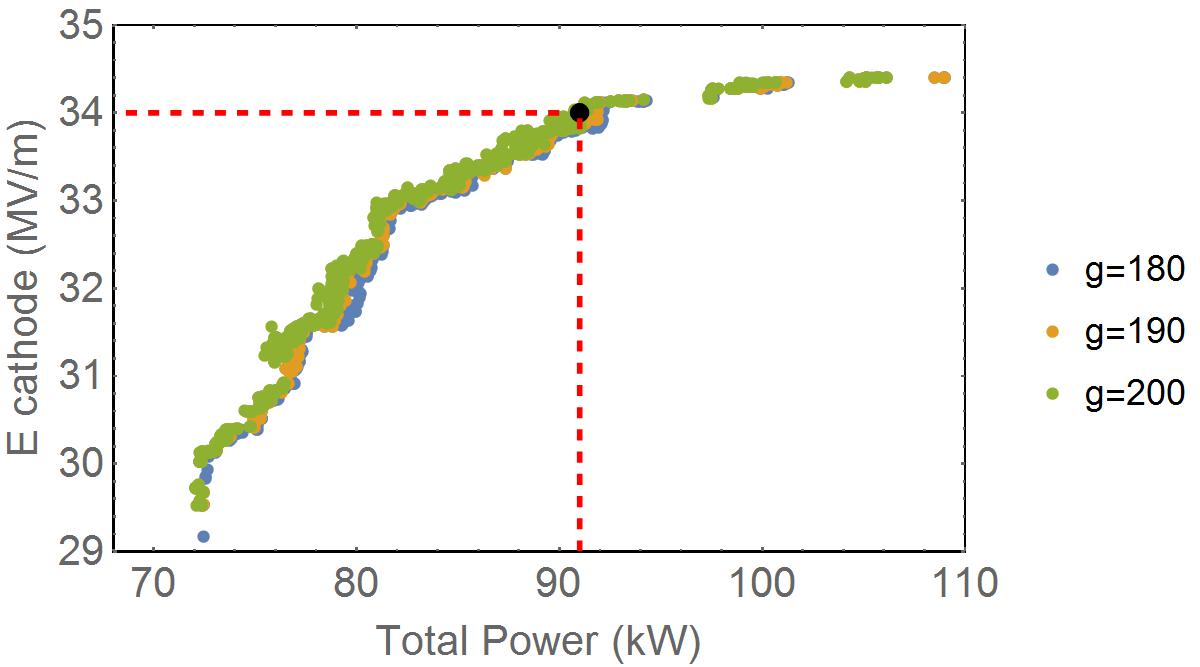}
\caption{\label{fig:gun_pareto} Pareto front of MOGA at $g=$180 (blue), 190 (yellow) and 200 (green) generation. The optimized solution is chosen from the  to have $E\textsubscript{cathode}\geq$ \SI{34}{MV/m} while the total power is minimum, shown as the black dot in the plot.}
\end{figure}

In MOGA, we chose a population $N=720$ and calculated up to $g=200$th generation. The Pareto fronts plotted at $g=180,\:190,\:200\:$ are shown in Figure~\ref{fig:gun_pareto}. Good convergence has been achieved at $g=200$. The Pareto front clearly shows the trade-off between a high $E\textsubscript{cathode}$ and a low $P\textsubscript{total}$. On the Pareto front of $g=200$, we chose a geometry with $E\textsubscript{cathode}\geq$ \SI{34}{MV/m} and the minimum $P\textsubscript{total}$ as the optimized solution, indicated as the dark dot in Figure~\ref{fig:gun_pareto}. 

\begin{figure}[hbt!]
\centering
\includegraphics[width=60mm]{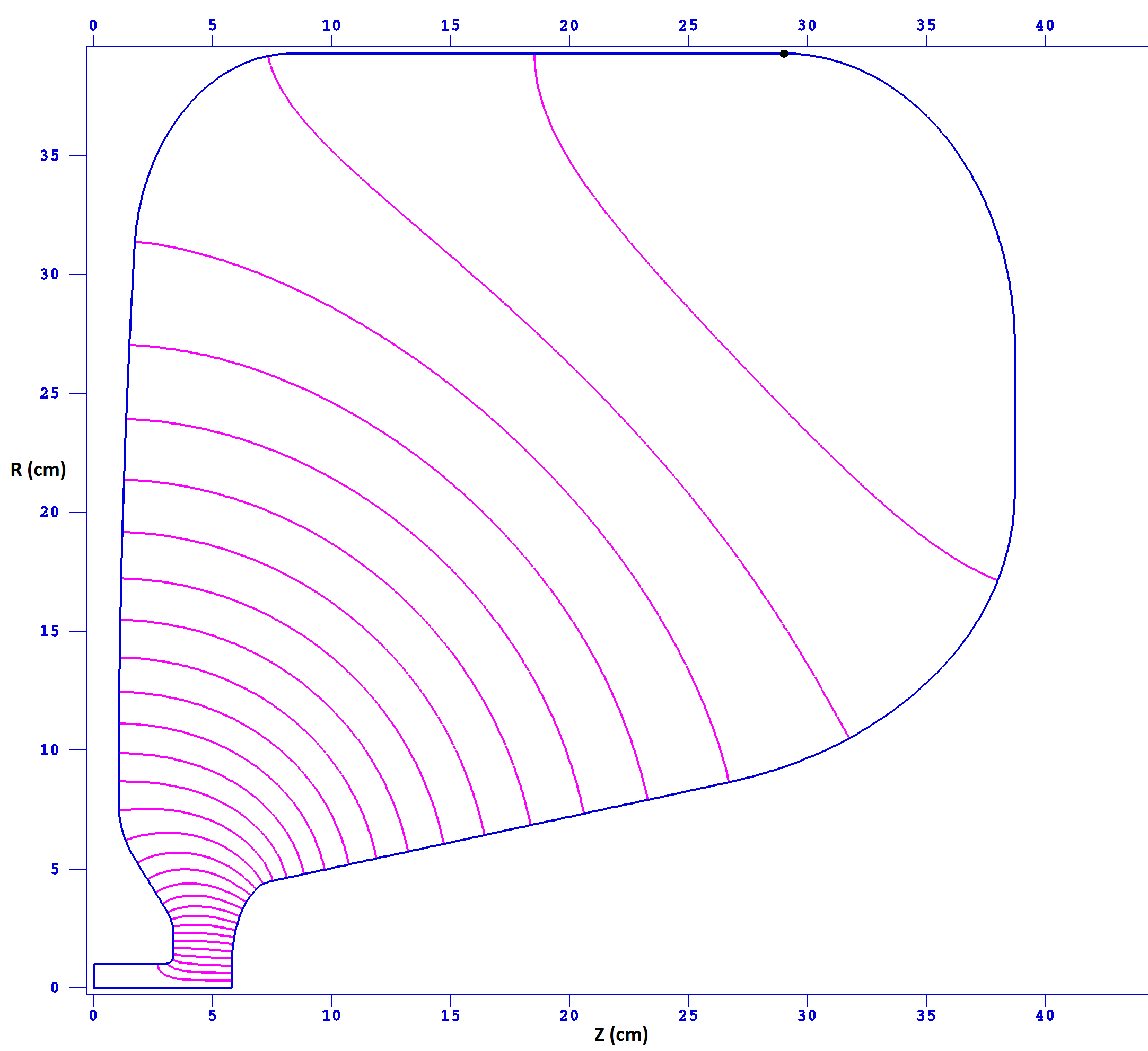}
\caption{\label{fig:gun_sf} SUPERFISH solution of optimized gun cell.}
\end{figure}

The SUPERFISH E field plot and main RF parameters of this optimized solution are shown in Figure~\ref{fig:gun_sf} and Table~\ref{sf_results}. Compared with the APEX gun, $E\textsubscript{cathode}$ of this design has increased significantly from 19.5 to \SI{34}{MV/m}. This improvement is mainly due to the decrease of the accelerating gap width $G$ from 4 to \SI{2.5}{cm}. Reducing the beampipe radius from 1.5 to \SI{1}{cm} also helps concentrating the E field along the beam axis. $P\textsubscript{total}$ is maintained at almost the same level as for APEX with \SI{90}{kW} input power. The output energy $V$ increases slightly from 750 to \SI{820}{kV}. Due to the large enhancement of $E\textsubscript{cathode}$, both $E\textsubscript{peak}$ and $PD\textsubscript{peak}$ are inevitably increased significantly compared to APEX. Still the ratio $E\textsubscript{peak}$/$E\textsubscript{cathode}$ is lower than APEX.

\begin{table*}
\caption{Main geometry and RF parameters of the optimized gun cell and 2\textsuperscript{nd} cell. APEX gun cavity parameters are also included as a reference. \label{sf_results} }
\begin{tabular}{lccc}\hline\hline
\textbf{Cavity parameters} & \textbf{Gun Cell} &\textbf{2\textsuperscript{nd} Cell} & \textbf{APEX}\\ \hline
Radius $R$ (cm) & 39.3 &39.1 & 36.0 \\
Length $L$ (cm) & 38.7 &36.0 & 35.0 \\
Accelerating gap $G$ width (cm) & 2.5 &4.6 & 4.0 \\
Beam pipe radius $r$ (cm) & 1.0 &1.0/1.5 & 1.5\\
RF frequency $f$ (MHz) & 162.5 & 162.5 & 185.7 \\ 
Total voltage $V$ (kV) & 820 &820 & 750 \\
Total power $P\textsubscript{total}$ (kW) & 90.7 & 85.4&88.5\\ 
Peak power density $PD\textsubscript{peak}$ (W/cm\textsuperscript{2}) & 32.1 &29.8 &22.8 \\
Cathode launching field $E\textsubscript{cathode}$ (MV/m) & 34.0 &NA  & 19.5\\
Peak surface field $E\textsubscript{peak}$ (MV/m) & 37.0 &24.7 &24.0 \\
$E\textsubscript{peak}$/$E\textsubscript{cathode}$ & 1.09 & NA &1.23 \\
\hline\hline
\end{tabular}
\end{table*}

\begin{figure}[hbt!]
\centering
\includegraphics[width=80mm]{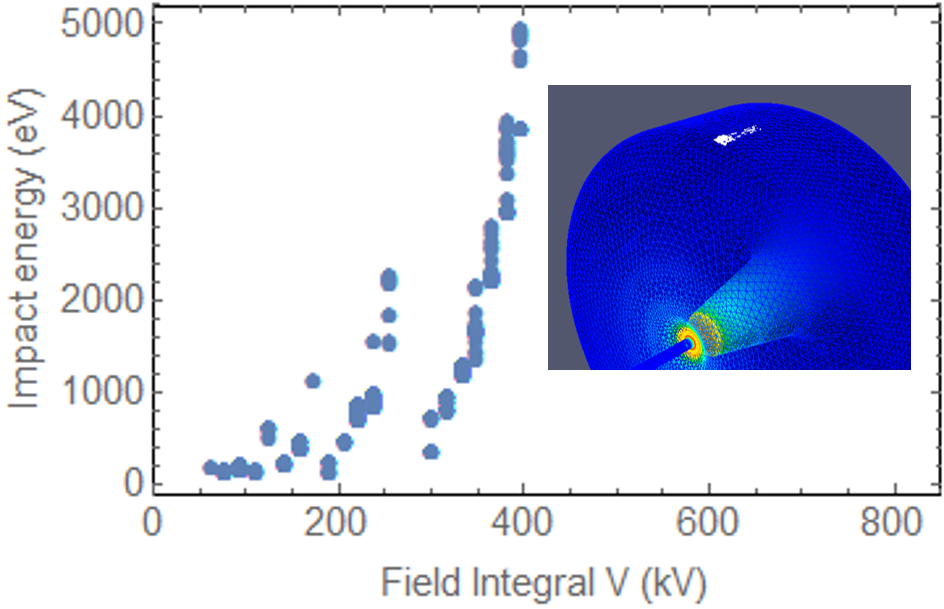}
\caption{\label{fig:gun_mp}MP impact energy spectrum of gun cell. No MP is observed at the cavity operation power level $V =$ \SI{820}{kV}. Some MP patterns are present when $V<$ \SI{400}{kV}. The MP in the cavity happens along the torus near the anode wall corner, shown as the white dots in the figure.}
\end{figure}

Since the cavity will be operated under Continuous Wave (CW) mode for high repetition operations, the MultiPacting (MP) resonance performance  needs to be examined. MP simulations are carried out with Track3P of ACE3P to identify the resonant motions that can produce hazardous MP. The MP impact energy spectrum is shown in Figure~\ref{fig:gun_mp}. No MP pattern is found at the operating power level. Similar to APEX, some MP patterns appear at low power range, mainly located around the outer corner on the anode wall. 

\subsection{The 2\textsuperscript{nd} Cell Design}

\begin{figure}[hbt!]
\centering
\includegraphics[width=70mm]{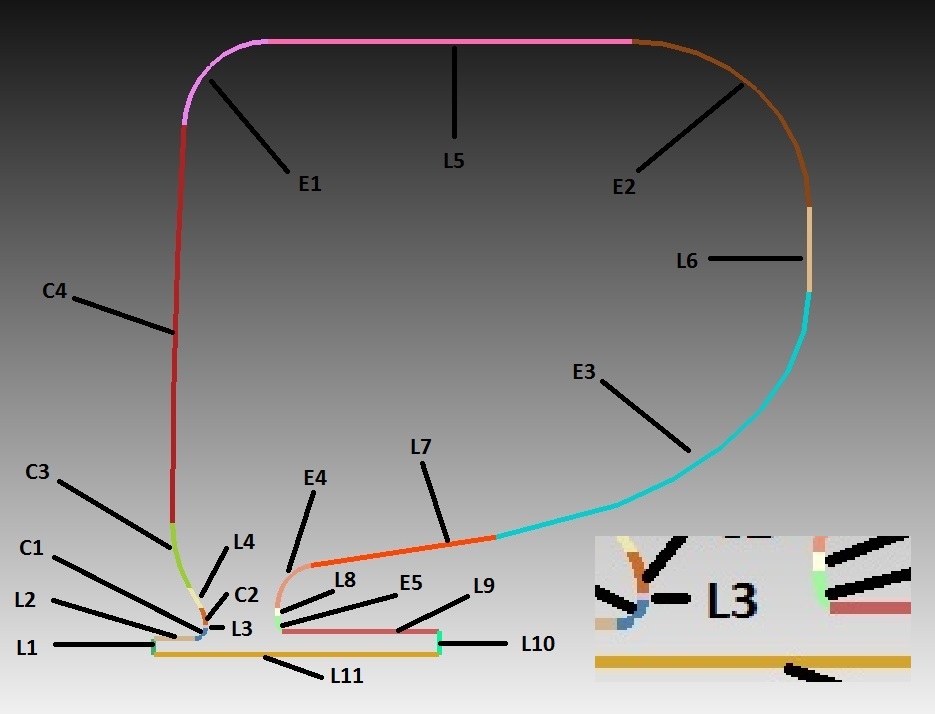}
\caption{\label{fig:2nd_cell_outline} Geometrical description of 2\textsuperscript{nd} cell. It consists eleven straight lines L1 to L9, four circular arcs C1 to C4 and five elliptical arcs E1 to E5. The zoom-in view of the gap region is at the bottom right}
\end{figure}

The 2\textsuperscript{nd} cell geometry is similar to the gun cell except there is no cathode plug. It is described by twenty segments, including eleven straight lines, four circular arcs and five elliptical arcs, as shown in Figure~\ref{fig:2nd_cell_outline}. Beampipe radii L1 and L10 are set at \SI{1}{cm} and \SI{1.5}{cm} respectively. Beampipe length L2 and L9 are set at \SI{10}{cm}. The remaining 16 segments are described by 21 independent parameters. 

The main function of the 2\textsuperscript{nd} cell is to provide further acceleration for electron beams. For a fixed total voltage, we choose the low peak surface E field and the low RF power loss as the design priorities. Other considerations include the RF frequency, the peak power density, the practical size, the connection to the gun cell and the space for installing the focusing solenoid. The beam dynamics simulation~\cite{chad-apex2-emittance} shows that the focusing solenoid should be placed close to the cathode to achieve good emittance compensation, thus the accelerating gap of the 2\textsuperscript{nd} cell cannot be too large. At the same time, the gap can neither be too small considering the restriction on the total RF power. The objectives and constraints in MOGA are listed in Table~\ref{moga_2nd_cell}.

\begin{table*}
\caption{MOGA optimization setting for 2\textsuperscript{nd} Cell design \label{moga_2nd_cell} }
\begin{tabular}{ll}
\hline\hline 
\textbf{Objectives}&\textbf{Constraints} \\ \hline
With total voltage $V=$ \SI{820}{kV}: & Accelerating gap $G$ plus chamfers on both ends $<$ \SI{5.7}{cm} \\ 
1) Minimize peak surface field  $E\textsubscript{peak}$ & Peak power density $PD\textsubscript{peak} <$ \SI{30}{W/cm\textsuperscript{2}} \\
2) Minimize total RF power $P_\textsubscript{total}$ & RF frequency $f=162.5 \pm$ \SI{3}{MHz} \\
& Cavity radius $R<$ \SI{39.3}{cm} \\ 
& Extrusion one anode side $K<$ \SI{1.5}{cm} \\
\hline\hline
\end{tabular}
\end{table*}

\begin{figure}[hbt!]
\centering
\includegraphics[width=80mm]{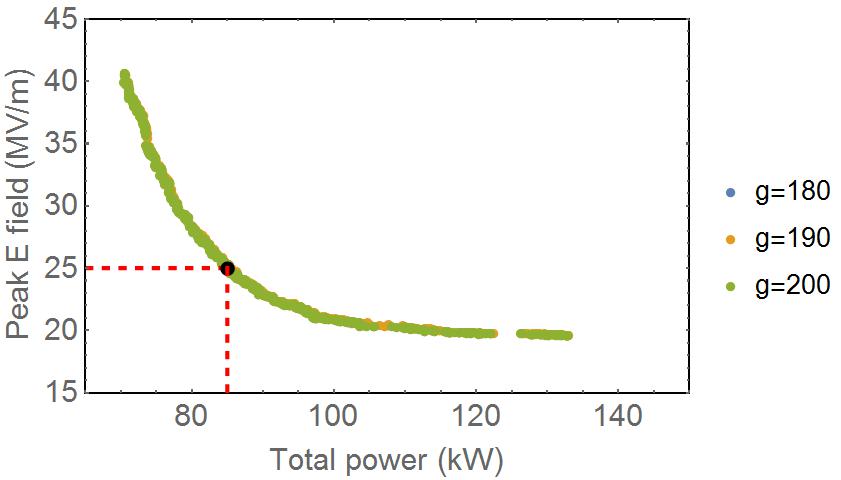}
\caption{\label{fig:2nd_pareto} Pareto front of MOGA at $g=$180 (blue), 190 (yellow) and 200 (green) generation. The optimized solution is chosen to have $E\textsubscript{peak}\leq$ \SI{25}{MV/m} while the total power is minimum, indicated as the black dot in the plot.}
\end{figure}

Same as the gun cell, we chose a population $N=720$ and carried out the calculation up to $g=200$ generation. Good convergence has been achieved at $g=200$, as shown in the Pareto front plot in Figure~\ref{fig:2nd_pareto}. The Pareto front shows a trade-off between a low $E\textsubscript{peak}$ and a low $P\textsubscript{total}$ as expected. On the Pareto front of $g=200$, we chose a geometry with $E\textsubscript{peak}\leq$ \SI{25}{MV/m} and the minimum $P\textsubscript{total}$ as the optimized solution, indicated as the dark dot in Figure~\ref{fig:2nd_pareto}.     

\begin{figure}[hbt!]
\centering
\includegraphics[width=60mm]{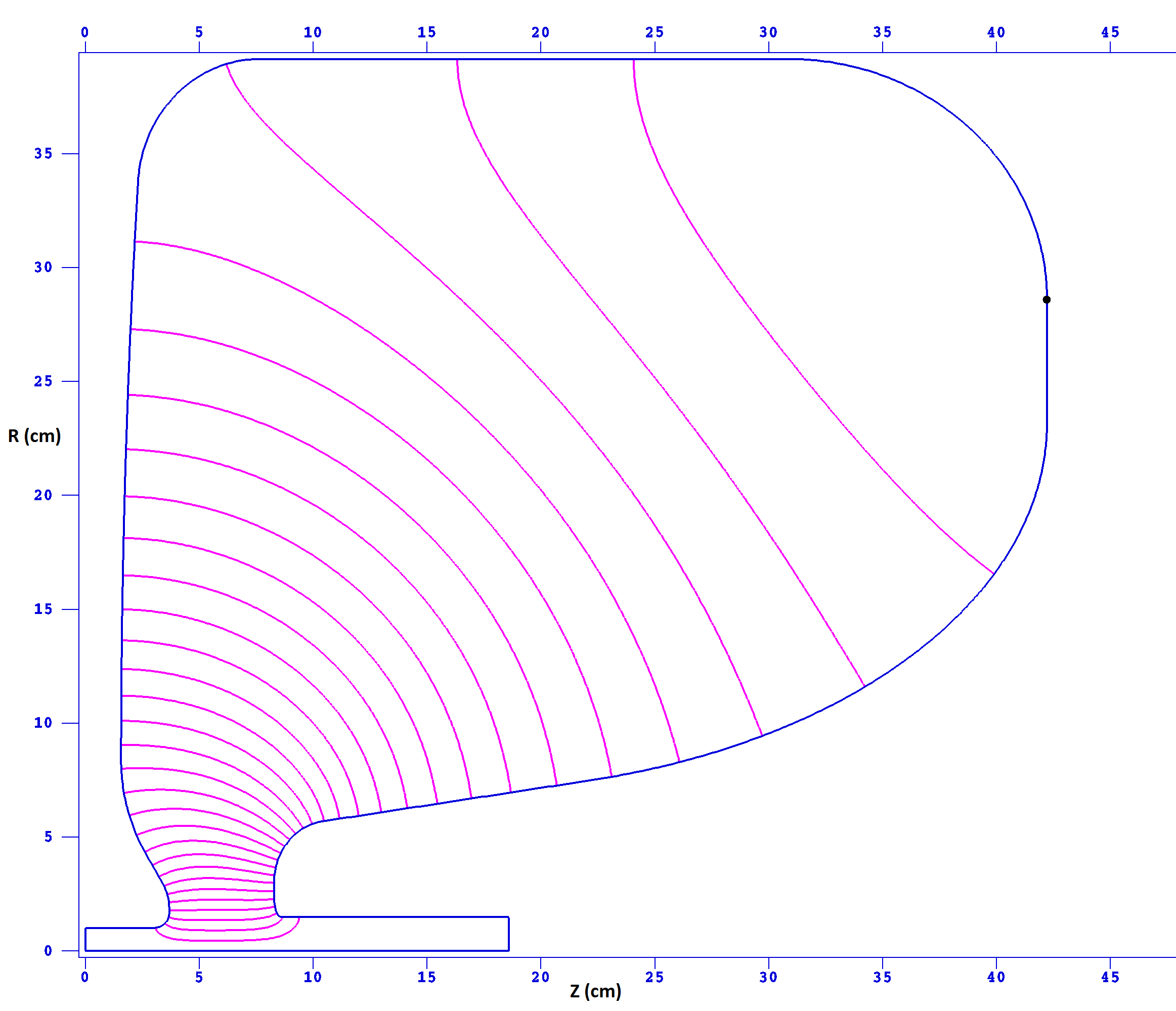}
\caption{\label{fig:2nd_sf} SUPERFISH solution of optimized 2\textsuperscript{nd} cell.}
\end{figure}

The SUPERFISH E field plot and the main RF parameters of this solution are shown in Figure~\ref{fig:2nd_sf} and Table~\ref{sf_results}. Compared with the gun cell, the gap width $G$ is increased to 4.6 cm to reduce $E\textsubscript{peak}$ and $PD\textsubscript{peak}$. The radial size is similar to the gun cell right below \SI{40}{cm}. The total RF power at \SI{85}{kW} is also similar to the gun cell.

\begin{figure}[hbt!]
\centering
\includegraphics[width=70mm]{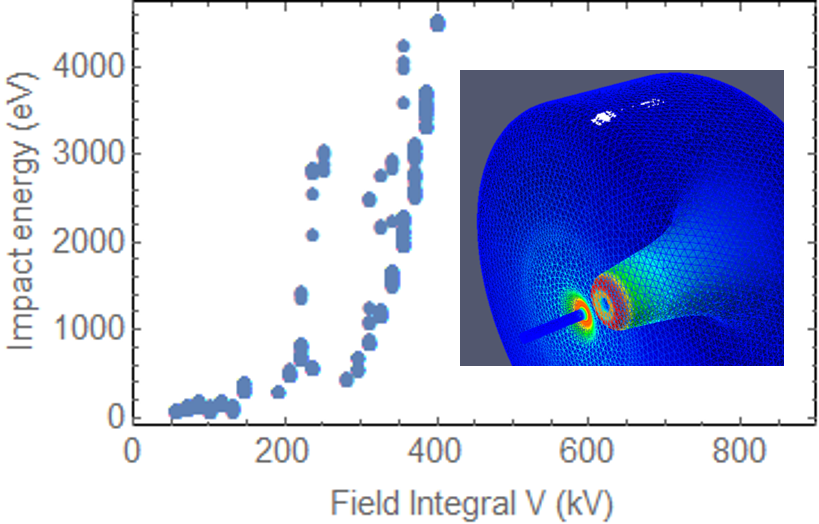}
\caption{\label{fig:2nd_mp}MP impact energy spectrum of 2\textsuperscript{nd} cell. The results are very similar to gun cell. No MP is observed at the cavity operation power level $V =$ \SI{820}{kV}. Some MP patterns are present when $V<$ \SI{400}{kV}. The MP in the cavity happens along the torus near the anode wall corner, shown as the white dots in the figure.}
\end{figure}

The MP simulation results by ACE3P is shown in Figure~\ref{fig:2nd_mp}. Same as the gun cell, no MP is found at the operational power level. Some MP patterns appear at the low power level, mainly located around the outer corner on the anode wall. 

\subsection{Two-cell Cavity Design}

\begin{figure}[hbt!]
\centering
\includegraphics[width=60mm]{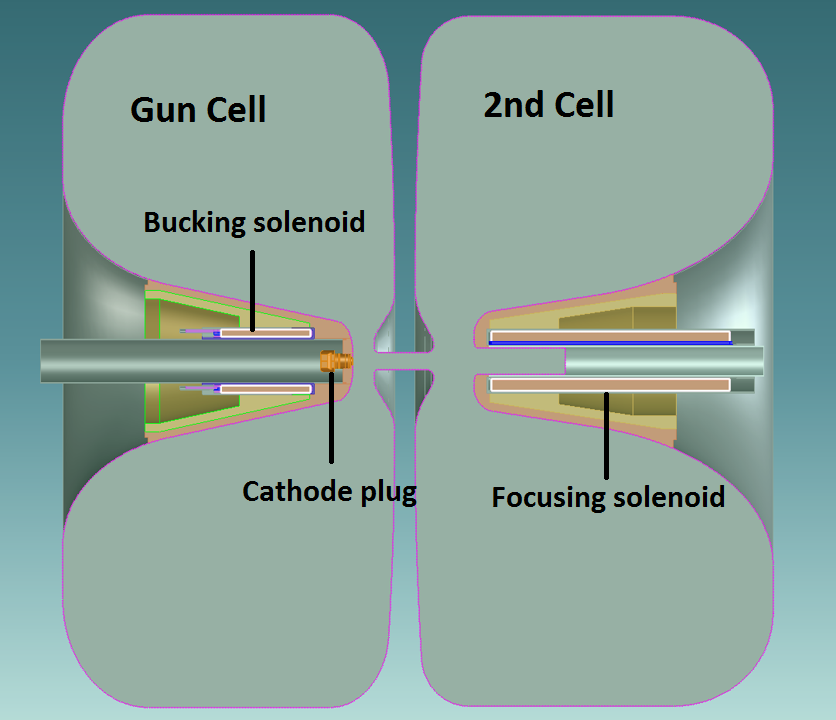}
\caption{\label{fig:two-cell-layout}The 2-cell layout of APEX2 gun cavity. }
\end{figure}

With the design of each cell done, they are put together to make the complete two-cell cavity, as shown in Figure~\ref{fig:two-cell-layout}. The distance between the cells is set large enough to prevent RF coupling but also small enough to minimize the beam size growth along the structure, which would lead to a consequent emittance increase at the solenoid due spherical aberrations. An injection beamline with this two-cell cavity has achieved a normalized emittance $\epsilon_{xn}(95\%)\approx$ 0.09 $\mu$m with \SI{100}{pC} charge and \SI{12.5}{A} peak current~\cite{chad-apex2-emittance}, which is a significant improvement over the state-of-art CW guns. 

\begin{figure}[hbt!]
\centering
\includegraphics[width=80mm]{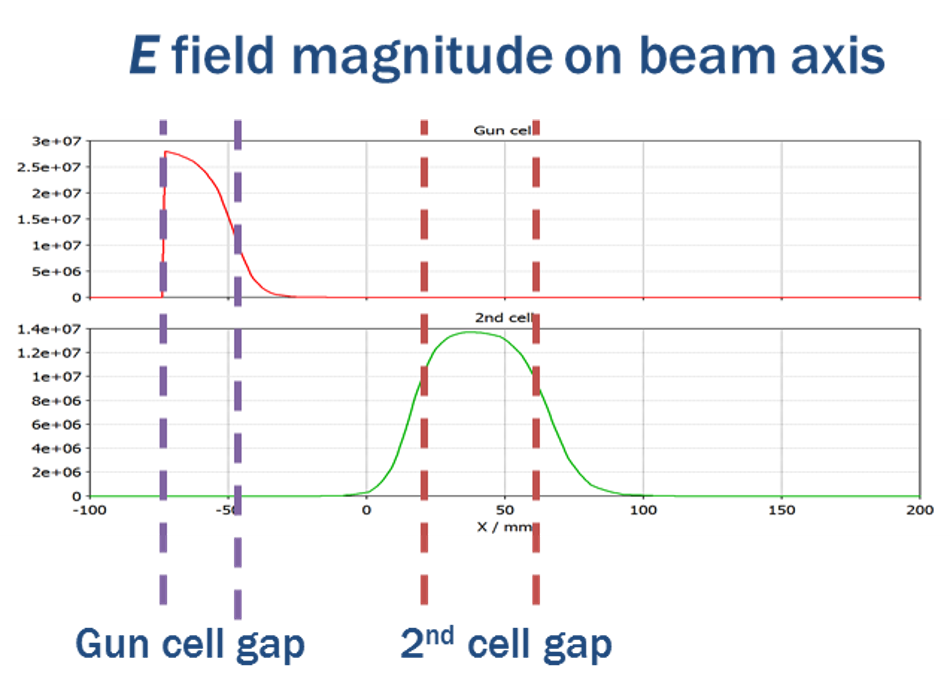}
\caption{\label{fig:2-cell-efield}The on-axis E field when exciting each cell individually. }
\end{figure}

The RF coupling between the two cells is extremely weak. The cutoff frequency of TE11 and TM01 mode of the $r=$ \SI{1}{cm} beampipe are:
\begin{equation}
\begin{split}
f\textsubscript{c\_TE11}&=\frac{c}{2\pi}\frac{1.841}{r}=8.79\: \textrm{GHz},\\
f\textsubscript{c\_TM01}&=\frac{c}{2\pi}\frac{2.405}{r}=11.48\: \textrm{GHz},
\end{split}
\end{equation}
both of which are much larger than the cavity operating frequency. The decay length of each mode:
\begin{equation}
\begin{split}
l\textsubscript{TE11}&\approx \frac{r}{1.841}=0.54 \: \textrm{cm},\\
l\textsubscript{TM01}&\approx \frac{r}{2.405}=0.42\: \textrm{cm},
\end{split}
\end{equation}
both of which are much smaller than the length of the beampipe between the two cells $d=$ \SI{5.68}{cm}. The plots of E field on axis when driving each cell are shown in Figure~\ref{fig:2-cell-efield}.

\section{Discussion}

\subsection{General Geometic Effects of the Re-entrant VHF Gun Cavities}
During the design of APEX2 gun cavity with MOGA, we found some general correlations between certain geometric parameters and the RF properties for such re-entrant VHF gun cavities.
\begin{enumerate}
\item{With fixed $V$, $E\textsubscript{cathode}$ is mainly determined by the gap width $G$.}
\item{$E\textsubscript{peak}$ is located at the edge of the cathode flat area. The smaller the cathode flat area, the lower the ratio $E\textsubscript{peak}/E\textsubscript{cathode}$ can be. }
\item{Cavities with larger radius $R$, smaller cathode cone angle $\theta 1$ and larger anode extrusion $K$, as shown in Figure~\ref{fig:geometry}, generally tend to have less power loss $P\textsubscript{total}$. Larger elliptical corner radius $R\textsubscript{ia}/R\textsubscript{ib}$ also helps reduce $P\textsubscript{total}$. }
\end{enumerate}

\subsection{A Preliminary \SI{216.7}{MHz} Gun Cavity Design}
Besides the frequency of \SI{162.5}{MHz}, we have also considered the option of \SI{216.7}{MHz}. From the RF design point of view, the main advantages of a \SI{216.7}{MHz} cavity are:
\begin{enumerate}
\item{The Kilpatrick limit~\cite{kilpatrick} is increased from \SI{13.6}{MV/m} to \SI{15.2}{MV/m}. Since the gradient of $E_\textsubscript{peak}$ of \SI{37}{MV/m} has never been demonstrated at VHF range with CW operation, at least to the authors' knowledge, operating at a higher frequency can reduce the risk of potential RF breakdown.}
\item{The higher frequency results in smaller cavity size. During the design of \SI{162.5}{MHz} cavity, we limit the cavity radius comparable to APEX, which compromises the reduction of the total power consumption. For the \SI{216.7}{MHz} cavity, the same radius restriction becomes relatively larger due to the shorter wavelength. Thus during the opimization, MOGA can explore the geometric parameter space more extensively for the better solutions.}
\end{enumerate}

\begin{figure}[hbt!]
\centering
\includegraphics[width=70mm]{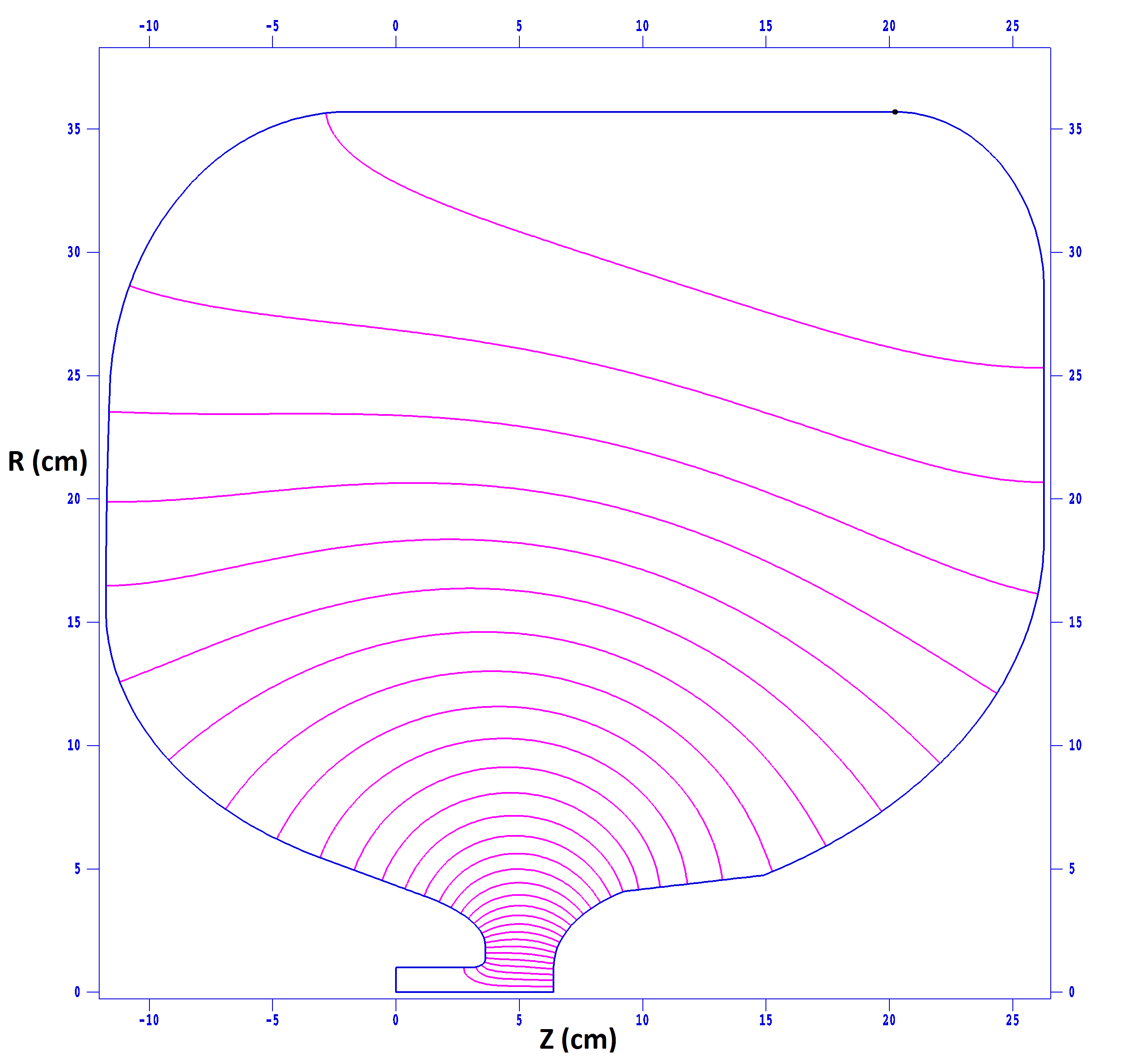}
\caption{\label{fig:217_gun} SUPERFISH solution of preliminary optimized \SI{216.7}{MHz} gun cell.}
\end{figure}

A preliminary \SI{216.7}{MHz} gun cell cavity design is shown in Figure~\ref{fig:217_gun}, with its main parameters listed in Table~\ref{217_parameters}. The result looks promising. Due to the reduced eigenmode wavelength and fixed cathode flat area, the ratio $E_{peak}/E\textsubscript{cathode}$ is decreased. With the same $V=820$ kV, both $P\textsubscript{total}$ and $PD\textsubscript{peak}$ are reduced significantly. $E\textsubscript{cathode}$ decreases to \SI{32}{MV/m}, given the same $E\textsubscript{peak}$ at \SI{37}{MV/m}, which likely can be restored back to \SI{34}{MV/m} by reducing the cathode flat area.  
 
\begin{table}
\centering
\caption{\label{217_parameters}Main geometry and RF parameters of the preliminary design of a \SI{216.7}{MHz} gun cell. \SI{162.5}{MHz} APEX2 gun cell is also listed in the third column as a reference.}
\begin{tabular}{lcc}
\hline\hline
\textbf{Cavity parameters} & \textbf{216.7 MHz}& \textbf{162.5 MHz}\\ \hline
 $R$ (cm) & 35.7 &39.5 \\
$L$ (cm) & 38.0 &38.7\\
 $G$ width (cm) & 2.7 &2.5\\
$r$ (cm) & 1.0 &1.0\\
$V$ (kV) & 820 & 820 \\
$P\textsubscript{total}$ (kW) & 55.8 & 90.7\\ 
$PD\textsubscript{peak}$ (W/cm\textsuperscript{2}) & 25.4 & 32.1 \\
 $E\textsubscript{cathode}$ (MV/m) & 32.1 & 34.0\\
$E\textsubscript{peak}$ (MV/m) & 37.0 & 37.0 \\
$E\textsubscript{peak}$/$E\textsubscript{cathode}$ & 1.15 & 1.09 \\
\hline\hline
\end{tabular}
\end{table}

\subsection{Future Development of MOGA-based RF Cavity Design Method}
The design of APEX2 gun cavity has shown the MOGA-based method is a very useful tool for RF cavity design. There are still several improvements to carry out in the future to make this method more efficient and applicable to more design cases:
\begin{enumerate}
\item{Replace SUPERFISH with a modern Linux-friendly 2D $EM$ field solver with more efficient calculation and more flexible geometry definition. Explore 3D $EM$ field solvers.}
\item{Explore other MOGAs beyond NSGA-II that can be particularly compatible with $EM$ field solvers.}
\item{Build the program portable to large scale computer cluster to fully utilize the capability of parallel computing. }
\end{enumerate}

\section{Conclusion}

The RF design of a \SI{162.5}{MHz} two-cell re-entrant VHF RF gun cavity has been presented in this paper. The optimization has led to a significant increase of $E\textsubscript{cathode}$ and $V$ compared to the previous generation of CW normal-conducting guns (19.5 to \SI{34}{MV/m} and \SI{750}{keV} to \SI{1.5}{MeV}). This improvement is likely to lead to a considerable enhancement of the injector beam brightness. 

A novel RF cavity design method has been developed and implemented in the design procedure. Combining the GA, the EM field solver and the parallel computing, this method proves to be a useful tool for the RF cavity design.

\section{Acknowledgments}
The authors would like to thank Dr. Houjun Qian at Deutsches Elektronen-Synchrotron, Germany and Dr. Changchun Sun at LBNL for the helpful discussions. This work is supported by Director of Science of the U.S. Department of Energy under Contract No. DE-AC02-05CH11231. The research used resources of the National Energy Research Scientific Computing Center, a U.S. Department of Energy Office of Science User Facility operated under Contract No. DE-AC02-05CH11231.

\bibliography{apex2}

\end{document}